\documentclass{rspublic}
\usepackage{graphicx}

\newcommand{\ea}{{\it et al. }}
\newcommand{\apj}{{\it Astrophys. J.}}
\newcommand{\apjs}{{\it Astrophys. J. Suppl. Ser.}}
\newcommand{\aj}{{\it Astron. J.}}
\newcommand{\mnras}{{\it Mon. Not. R. Astron. Soc.}}
\newcommand{\aanda}{{\it Astron. Astrophys.}}
\newcommand{\pasp}{{\it Publ. Astron. Soc. Pac.}}
\newcommand{\ptra}{{\it Phil. Trans. R. Soc. A}}

\begin{document}

\title[Cluster formation: simulations]{The physics and modes of star cluster
formation: simulations}

\author[C. Clarke]{Cathie Clarke}

\affiliation{Institute of Astronomy, University of Cambridge,
Madingley Road, Cambridge CB3 0HA, UK}

\label{firstpage}

\maketitle

\begin{abstract}{{\bf hydrodynamics; stars: formation; stars: mass
function; open clusters and associations: general}} We review progress
in numerical simulations of star cluster formation.  These simulations
involve the bottom-up assembly of clusters through hierarchical
mergers, which produces a fractal stellar distribution at young ($\sim
0.5$\,Myr) ages. The resulting clusters are predicted to be mildly
aspherical and highly mass-segregated, except in the immediate
aftermath of mergers. The upper initial mass function within
individual clusters is generally somewhat flatter than for the
aggregate population. Recent work has begun to clarify the factors
that control the mean stellar mass in a star-forming cloud and also
the efficiency of star formation. The former is sensitive to the
thermal properties of the gas while the latter depends both on the
magnetic field and the initial degree of gravitational boundedness of
the natal cloud. Unmagnetized clouds that are initially bound undergo
rapid collapse, which is difficult to reverse by ionization feedback
or stellar winds.
\end{abstract}

\section{Introduction}

Hydrodynamic simulations of star cluster formation are in their
infancy. This is hardly surprising, since only in the last decade has
it been possible to move beyond modelling the formation of single
stars and binaries. Even with current high-performance-computing
facilities, it is still unfeasible to undertake cluster simulations
that can follow the collapse of individual stars down to stellar
densities (cf. Bate 1998).  Thus, `star formation' in the simulations
actually means the accumulation of gravitationally bound gas within
`sink particles' of specified radius.  In general, the most ambitious
simulations (in terms of numbers of stars formed and, hence, the
ability to track hierarchical cluster assembly) are those that
necessarily use rather large sink radii. This suppresses the proper
modelling of close-binary formation (see Goodwin 2010) and compromises
the modelling of stellar dynamical interactions. Other shortcomings
include the fact that most simulations to date omit magnetic fields
and model the gas using a simple barotropic equation of state.

These shortcomings should not detract from the achievement of being
able to model the hydrodynamics of cluster formation at all. First, it
should be stressed that there are a number of ongoing efforts to add
new physical ingredients to the simulations, including magnetic
fields, radiative transfer and the effects of thermal and mechanical
feedback. Second, however, it should be borne in mind that even in
their simplest, `vanilla' forms, these simulations represent an
enormous advance on attempts to model young clusters as an $N$-body
system plus background potential (e.g., Geyer \& Burkert 2001; Boily
\& Kroupa 2003). 

Although such simulations can be valuable in studying the later stages
of cluster dispersal, once star formation has ceased (Goodwin \&
Bastian 2006), they are no substitute during the early, embedded phase
of cluster evolution for hydrodynamical simulations that can capture
the complex interplay between fragmentation, accretion, infall and
stellar dynamics. This review concentrates on predictions of those
`turbulent-fragmentation' simulations that have been most thoroughly
analysed to date. The discussion will, therefore, focus on observables
(e.g., mass segregation, mass functions, clustering statistics,
cluster morphology) and should therefore be read in conjunction with
Lada (2010), who reports on the current observational status in these
areas. This is followed by a preliminary survey of the results to date
from simulations that include a more complete suite of input physics.
Finally, we try and link the insights gained from such simulations to
issues of star formation on a Galactic scale.

Before embarking on this description, it is worth stressing that the
simulations described here relate to the very earliest (deeply
embedded) phases of cluster formation, typically covering about the
first 0.5 Myr following collapse of the parent cloud. Clustering
statistics are still evolving at the end of these simulations.
Consequently, questions of ultimate cluster survival have to be
followed over considerably longer timescales [see the discussion of
cluster `infant mortality' by de Grijs (2010)].

\section{An overview of cluster-formation simulations}

\subsection{Setting up  of `turbulence' in the cluster-formation context}

It is now well-established that molecular clouds evidence supersonic
internal motions whose energy is similar to the gravitational
potential energy of the cloud (Larson 1981; Heyer \ea 2001; Heyer \&
Brunt 2004). Thus, molecular cloud complexes are likely to contain
both regions that are somewhat bound and somewhat unbound and this is
an important factor in determining the star-formation efficiency and
merger history of clusters locally (Clark \& Bonnell 2004; Clark \ea
2008). The observed `size--linewidth' relation in molecular clouds
(Larson 1981) implies a velocity power spectrum with index $P(k)
\propto k^{-4}$. This similarity to the spectra of Kolmogorov or
Burgers turbulence has encouraged an interpretation in terms of a
cascade of energy from some large scale where it is being continuously
injected. Grid-based simulations (e.g., Kritsuk \ea 2007; Lemaster \&
Stone 2008; Kitsionas \ea 2008) are well suited to modelling the
resulting turbulent density fields.

Such simulations, however, do not include self-gravity and so the
stellar mass functions derived from such simulations are obtained a
posteriori by identifying what would be Jeans-unstable fragments in
the density field (Padoan \ea 1997). Naturally, such (zero-gravity)
calculations cannot follow cluster formation and evolution. We
therefore concentrate here on self-gravitating calculations, which
(for the large-scale calculations discussed here) means Lagrangian
(smoothed-particle hydrodynamics; henceforth SPH) calculations [see
Chapman \ea (1992) and Klessen \ea (1998) for the earliest examples of
cluster-formation simulations]. These treat the velocity field in two
ways. Those of Klessen and collaborators (e.g., Schmeja \& Klessen
2004) impose continuous forcing of the velocity field at some range of
input scales, so that (as in the calculations of Padoan and
collaborators) a quasi-stationary turbulent cascade is set up.

Bonnell, Bate and co-workers instead introduce internal motions in the
cloud as a one-off initial input. Dissipation of these motions means
that star formation and cluster merging proceeds over one or two
global free-fall timescales, since there is no continuous injection of
kinetic energy to support the cloud. In this case, the velocity field
of the gas over a free-fall time is just inherited from the initial
conditions and therefore for the velocity field to match the observed
size--linewidth relation, they require a particular input velocity
power spectrum, i.e., $P(k) \propto k^{-4}$. It can justifiably be
argued that such a situation is not one of true turbulence, as it does
not reflect an equilibrium cascade of energy from large to small
scales. Nevertheless, we will use the term `turbulent' in the loose
way that has become common currency in the field. The reader should
bear in mind that this merely denotes a situation where the gas is
(either initially or continuously) subject to supersonic internal
motions.

\subsection{Structure formation and the implementation of star formation}

Supersonic turbulent motions lead to the formation of shocked layers
which, under the action of self-gravity, fragment into filaments and
then into bound objects. The computational expense of following the
hydrodynamic collapse of extremely dense gas is avoided through the
implementation of `sink particles' (Bate \ea 1995), which interact
gravitationally with the rest of the calculation and can also continue
to accrete bound gas that enters within the sink radius. The exact
criterion for sink-particle creation changes slightly between
calculations, but usually involves the requirement that $\sim 100$
bound SPH particles are contained within the sink at some minimum
threshold density (whose value varies between different simulations in
the range $\sim 10^{-13}-10^{-11}$ g cm$^{-3}$). We will henceforth
describe sink particles as `stars'; however, in the largest-scale
simulations, the sink radius is as large as 50 AU (Bonnell \ea 2003,
2004), so that modelling of both discs and close-binary companions is
thus impossible in these simulations.

\subsection{Evolution following the onset of star formation}

Stars typically form first in the dense gas at the intersection of
collapsing filaments (see figure 1).  Gas and stars forming along the
filaments' lengths fall in towards the local potential minimum and a
protocluster is then established (e.g., Klessen \& Burkert 2000, 2001;
Bate \ea 2003; Bonnell \ea 2003).  This is a small-scale system
initially and, therefore, few-body interactions are particularly
important for the first stars to form, especially in simulations with
small sink radii (Bate \ea 2003) which can model interactions with
moderately hard binaries (separations of a few AU).  The infalling gas
possesses angular momentum because of the vorticity of the initial
velocity field, and where the cluster centre is dominated by a single
massive star, it frequently acquires a large-scale disc (see lower
panels in figure 1).  This then fragments copiously, adding new stars
to the cluster in addition to those that arrive entrained in the
infalling gas.  [Note that this very copious fragmentation of
circumstellar discs is an artefact of the equation of state used in
most such calculations: see Bate (2009{\it b}) and \S4$\,a$.] The
initial mass of sink particles is very small ($\sim 10^{-3}$
M$_\odot$). They rapidly accrete the core of bound gas in their
vicinity but then, depending on their local environment, may also
continue to accrete significantly over much longer periods. This is
particularly true of the most massive stars, which rapidly segregate
to the gas-rich protocluster core.

\begin{figure}
\begin{center}
\includegraphics[width=0.75\columnwidth]{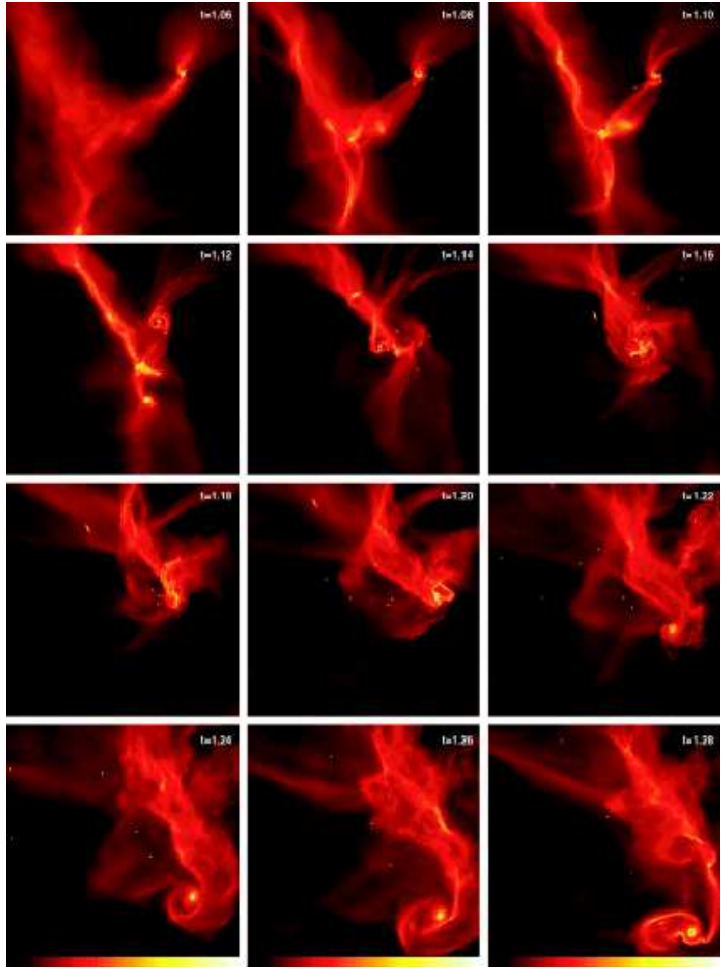}
\end{center}
\caption{Evolutionary sequence for cluster formation (Bate \ea 2003).}
\end{figure}

This differential acquisition of mass according to a star's current
mass is known as `competitive accretion' and was originally understood
in an idealized way in terms of Bondi--Hoyle accretion, whereby the
accretion cross section of a star scales quadratically with its mass
(Bondi \& Hoyle 1944). This accretion law causes a narrow range of
initial masses to be mapped onto a power-law distribution of final
masses, in which the fraction of stars with mass $M$ in a given linear
mass interval scales as $M^{-2}$ (Zinnecker 1982), suggestively close
to the Salpeter slope of $M^{-2.35}$ (Salpeter 1955).  Subsequently,
Bonnell \ea (2001) analysed competitive accretion in the context of
`plum pudding' models, in which stellar-mass points are embedded in a
smooth collapsing gas sphere. In the outer cluster, the stars almost
co-move with the gas and, hence, the relevant cross section is set by
the star's tidal radius rather than its (much larger) Bondi--Hoyle
radius. In this limit, the resulting IMF is predicted to be flatter
($\propto M^{-1.5}$) and Bonnell \ea ascribed the IMF in the
simulations (flatter at low mass and steepening to a Salpeter-like
high-mass tail) to competitive accretion in the tidally limited and
Bondi--Hoyle regimes, respectively.  It is remarkable, given the much
more complex dynamical histories of stars in the
turbulent-fragmentation calculations compared with the simple radial
infall in plum pudding models, combined with the extremely
inhomogeneous density fields in the turbulent case, that
turbulent-fragmentation calculations also produce an IMF of this form
(e.g., Bate \ea 2003; Bonnell \ea 2006; see figure 4).

Another aspect of competitive accretion is that the original
provenance of accreted material depends on a star's mass.  A low-mass
star, which remains of low mass (perhaps due to a dynamical
interaction that ejects it from its natal protocluster) just accretes
the overdense gas surrounding the sink particle at the time of its
creation. Stars that end up with high masses have protracted accretion
histories. They acquire mass from a variety of locations in the
initial cloud (Bonnell \ea 2004).

\subsection{The evolution of clustering: a bottom-up process}

Clustering in the simulations is a natural outcome of the highly
inhomogeneous density fields in turbulent clouds which are amplified
by gravitational collapse. It is a bottom-up process: the basic unit
of cluster formation is a small-$N$ group which successively merges
with larger entities. This requires a qualitative shift in our view of
cluster formation, away from traditional (monolithic-collapse) models.

\begin{figure}
\begin{center}
\includegraphics[width=0.75\columnwidth]{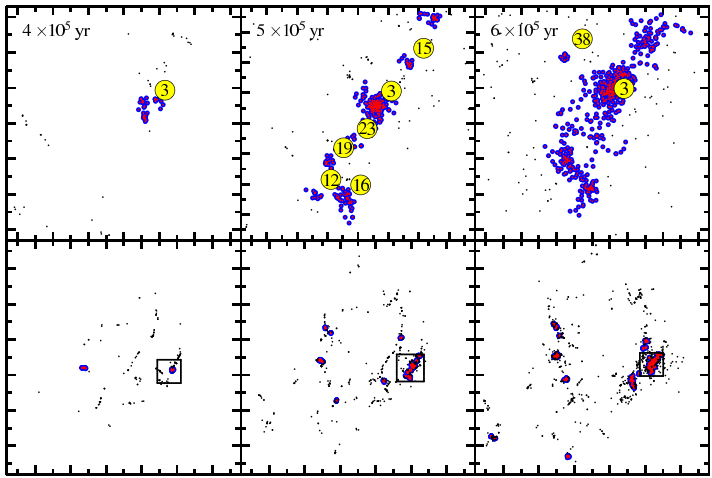}
\end{center}
\caption{\label{snapshot_times}\label{snapshot_large} Evolution of
clustering in the simulation of Bonnell \ea (2008). The upper panel
(scale $0.6 \times 0.6$ pc$^2$) is an enlargement of the boxed region
in the lower panel (scale $6 \times 6$ pc$^2$). (From Maschberger \ea
2009).}
\end{figure}

Figure 2 depicts the evolution of the stellar density distributions in
the simulation of Bonnell \ea (2008), while figure 3 represents the
corresponding merger tree (Maschberger \ea 2009). Clusters are
extracted from the projected stellar distributions using a minimum
spanning tree (Cartwright \& Whitworth 2004). The lower panel of
figure 2 illustrates the global evolution, wherein the left-hand part
of the simulation (which was initially mildly unbound) forms a number
of relatively isolated clusters and a pronounced field population
(small dots), while the boxed region (on which the upper panel zooms
in) undergoes a runaway merger to a single cluster.

\begin{figure}
\begin{center}
\includegraphics[width=0.6\columnwidth]{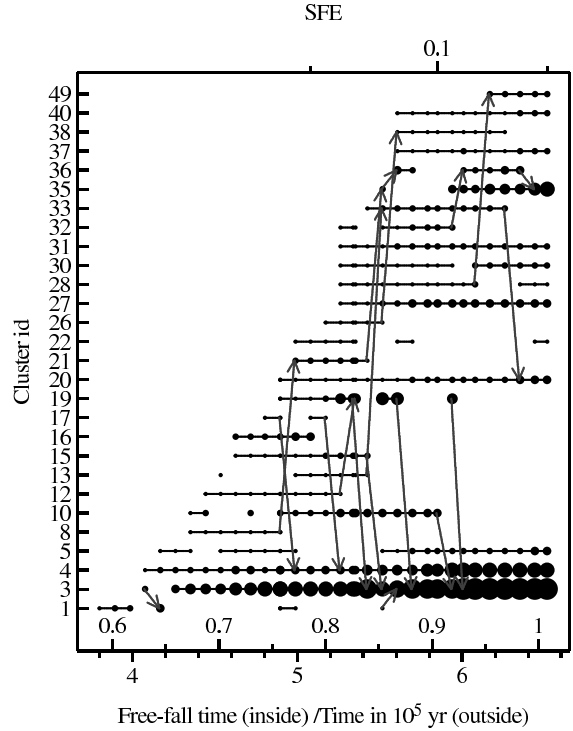}
\end{center}
\caption{Cluster merger tree from Maschberger \ea (2009) for the
simulation shown in figure 2.  The symbol size denotes the richness of
the cluster.}
\end{figure}

\section{Observational predictions of `vanilla' calculations}

We now examine the detailed results of what can be termed `vanilla'
calculations, i.e., those including the minimal subset of physical
effects that are required for turbulent fragmentation to occur.  Such
calculations are Lagrangian (SPH) and involve imposition of supersonic
internal motions. These simplest calculations omit magnetic fields and
apply a barotropic equation of state, which is a parameterization of
the results of spherically symmetric collapse calculations modelled
with frequency-dependent radiative transfer (Masanuga \& Inutsuka
2000). Here, we focus on large-scale calculations that model the
formation of at least hundreds of stars, so that we can analyse the
history of cluster assembly and the relationship to the IMF. Readers
with an interest in detailed predictions of the multiplicity
characteristics of smaller-scale cores are referred to Delgado--Donate
\ea (2003, 2004{\it a,b}), Goodwin \ea (2004{\it a,b}, 2006) and
Offner \ea (2009).

The calculations we discuss here are (i) those of Klessen and
collaborators (e.g., Klessen 2001; Schmeja \& Klessen 2004), which
employ periodic-box boundary conditions and a continual forcing of the
velocity field over a prescribed range of spatial scales. Both effects
counteract the global collapse that is seen in the other simulations,
although there is local collapse of regions where the ratio of
gravitational to kinetic energy is large; (ii) those of Bate \ea (Bate
\ea 2003; Bate 2009{\it c}) which focus on relatively small-scale
systems ($\sim 50$ M$_\odot$) to attain a good mass resolution and
small sink radius (note that Bate 2009{\it a} most recently extended
such high-resolution studies to larger clouds of 500 M$_\odot$). These
calculations are particularly well suited to the study of the low-mass
end of the IMF and best capture the formation and dynamical
interactions of close binaries; (iii) larger-scale simulations which
sacrifice mass and spatial resolution to be able to model a
statistically large ensemble of stars. These calculations are thus
optimal for studying the upper end of the IMF and the hierarchical
merging of clusters.

Much of what follows is based on the analysis of Maschberger \ea
(2009) of a $10^3$ M$_\odot$ simulation (reported in Bonnell \ea 2003,
2004) and a similar $10^4$ M$_\odot$ simulation (reported in Bonnell
\ea 2008).  However, we caution that it may be impossible to sacrifice
resolution on small scales without affecting the behaviour of the
simulation on larger scales.  The large sink radius in these
calculations (50 AU) suppresses the role of dynamical interactions
with even moderately hard binaries (with separations of tens of AU)
and therefore dynamical ejections play a less important role in these
simulations than in higher-resolution calculations.  Higher-resolution
`vanilla' calculations are, however, not necessarily more realistic
since, in contrast to the larger-scale simulations, they probably {\it
over\,}predict the role of close dynamical interactions because they
overproduce close companions (see \S4$\,a$).

\subsection{The formation of brown dwarfs}

The low-mass end of the IMF is populated by objects that form in
high-density gas (where the Jeans mass is relatively small) and whose
subsequent dynamical history does not cause them to continue to
accrete at a high rate. In practice, the high-density regions of the
simulation are either dense filaments or circumstellar discs (Bate \ea
2002). As noted above, the profuse fragmentation of circumstellar
discs in the simulations of Bate \ea (2002, 2003, 2009{\it a})
overproduces brown dwarfs, yielding a brown-dwarf-to-star number ratio
of at least $3:2$ compared with the observed estimate of $1:3$
(Andersen \ea 2008). This shortcoming in the barotropic calculations
has, however, been largely remedied by the inclusion of magnetic
fields and radiative feedback (see \S$\,a$).

The requirement of low accretion rates (such that brown dwarfs remain
brown dwarfs) is achieved in two ways in the simulations. In the
high-resolution simulations, the principal mechanism is through
dynamical ejections from regions of dense gas (Reipurth \& Clarke
2001). In low-resolution simulations, which suppress such effects,
another mechanism becomes apparent whereby brown dwarfs form in dense,
infalling filaments: here it is the tidal shear of such filaments as
they fall into the cluster potential that draws gas away from
collapsing condensations and thus ensures a low final mass (Bonnell
\ea 2008).

It should be stressed that, in contrast to the early speculations of
Reipurth \& Clarke (2001), the process of dynamical ejection does not
impart the resulting brown-dwarf population with a higher velocity
dispersion than the higher-mass stars. A clear observational hallmark
of the importance of dynamical interactions is that binary stars in
the simulations on average have a lower velocity dispersion than
single stars (Bate \ea 2009{\it a}; see also Delgado--Donate \ea
2003).

\subsection{The `knee' of the IMF}

\begin{figure}
\begin{center}
\includegraphics[angle=-90,width=0.8\columnwidth]{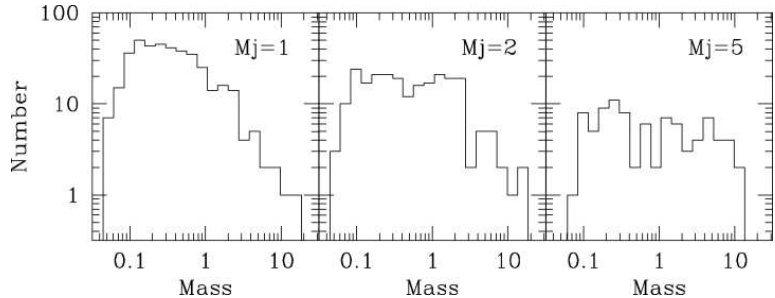}
\end{center}
\caption{The IMF produced in isothermal simulations of Bonnell \ea
  (2006) for three different values of the initial Jeans mass
  ($M_\mathrm{J}$, in units of M$_\odot$). Note that the `knee' of the
  IMF simply tracks $M_\mathrm{J}$.}
\end{figure} 

The IMF shown in figure 4 has the property that its power-law slope
(such that the fraction of stars in the mass range $M$ to
$M+\mathrm{d}M$ scales as $M^{-\alpha}$) makes a transition from
$\alpha < 2$ to $\alpha > 2$ at a mass scale of $\sim 1$
M$_\odot$. Since the mean stellar mass of a power-law distribution is
dominated by the upper (lower) mass limits for $\alpha < (>) 2$, we
see that such a rollover imparts the IMF with a characteristic stellar
mass at around this mass scale. The observed IMF (Kroupa \ea 1993)
shows similar behaviour and raises the issue of the origin of this
mass scale.

In the first generation of simulations (Bate \& Bonnell 2005), the
characteristic mass scaled simply with the average Jeans mass in the
{\it initial cloud}.  Padoan \ea (1997) (see also Clarke \& Bromm
2003) had previously argued that the mass scale is instead set by the
Jeans mass in the {\it shocked gas} and would therefore depend on the
Mach number of the turbulence. Although the Mach number undoubtedly
influences the density distribution attained in the simulations, the
full hydrodynamical calculations demonstrate that it has little effect
on the resulting IMF, probably because it is inappropriate to apply
the spherical Jeans criterion to shocked layers (Elmegreen \&
Elmegreen 1978; Lubow \& Pringle 1993; Clarke 1999).

There is, however, a glaring problem with any theory which links the
characteristic mass scale of the IMF with the gross properties of the
parent cloud, i.e., in which the characteristic stellar mass varies as
$T^{1.5}/\bar \rho^{1/2}$.  The order-of-magnitude variations in the
temperature ($T$) of molecular clouds, together with the difficulty of
defining a mean density ($\bar \rho$) in a fractally organized
interstellar medium (ISM), means that one would expect considerable
variation in the IMF knee in different star-forming regions. Instead,
this is found to be surprisingly invariant in all well-studied regions
(Kroupa 2002; Muench \ea 2002; Chabrier 2003). Clearly, therefore, we
need a theory which breaks the dependence of characteristic mass scale
on gross cloud properties and which instead invokes a physical process
to set a characteristic density and temperature.

To date, there are two (distinct) modifications of the simplest
barotropic equation of state used in the early calculations that can
break the mapping between mean cloud properties and characteristic
stellar mass scale.  Bonnell \ea (2006) replaced the strictly
isothermal equation of state (at densities below $10^{-13}$ g
cm$^{-3}$) by one which switched between mild cooling to mild heating
at a density of $3 \times 10^{-18}$ g cm$^{-3}$.  This prescription is
motivated by the arguments contained in Larson (2005), who identifies
the change in barotropic index with the onset of efficient dust--gas
coupling at this density (see also Whitworth \ea 1998).  With this
equation of state, Bonnell \ea were then able to produce
indistinguishable IMFs in simulations with quite different
parent-cloud parameters. 

We stress that these results refer to the {\it stellar} IMFs produced
by the simulations: although there is considerable interest in the
literature in relating this to the observed {\it core}-mass function
(CMF; Motte \ea 1998; Lada \ea 2008; Myers 2008), the analysis of the
simulations using a variety of {\sc clumpfind}-style algorithms (Smith
\ea 2008) yields CMFs for which the `knee' is a function of the
details of the algorithm employed.  Indeed, different observational
studies do not always agree on the identity of individual clumps in a
given region (Johnstone \ea 2000; Motte \ea 1998). Thus, although it
is important for simulations to go beyond the reproduction of the
stellar IMF and to also produce the correct structures in the ISM, it
is difficult to establish, pending wider agreement on clump-extraction
algorithms, whether or not this is the case.

Alternatively, the recent radiative-transfer simulations of Bate
(2009{\it b}) also disconnect the IMF and the mean Jeans mass of the
parent cloud. In this case, Bate argues that the characteristic mass
scale is set not by a particular cooling regime, but by the fact that
feedback from low-mass star formation produces a rather narrow spread
in the Jeans mass of irradiated gas (see \S4$\,a$).

\subsection{The high-mass tail of the IMF}

As noted above, the IMF in the simulations above the knee rolls over
and approaches a power law in the high-mass regime. Obviously, this is
much better sampled in the high-mass simulations. The simulations of
Bonnell \ea (2003, 2008) are particularly suitable in this regard as
at the end of the simulation they produce $\sim 100$ and $\sim 1000$
stars in the high-mass tail, respectively. The slope of this high-mass
tail is $\alpha = 2 \pm 0.2$, with some mild evolution towards a
flatter slope with time in regions where ongoing merging allows
massive stars to continue to accrete vigorously.

A notable result is that the slope of the upper IMF is flatter
(smaller $\alpha$) within individual clusters than for the whole
(cluster + field) population (figure 5).  This is partly a consequence
of mass segregation (see \S3$\,e$), since the more massive stars are
more likely located within clusters.  We see indeed in figure 5 that
the IMF for all stars in clusters is flatter than the IMF of the whole
(cluster + field) population, since the stellar composition of the two
samples is not identical.

\begin{figure}
\begin{center}
{\includegraphics[width=0.65\columnwidth]{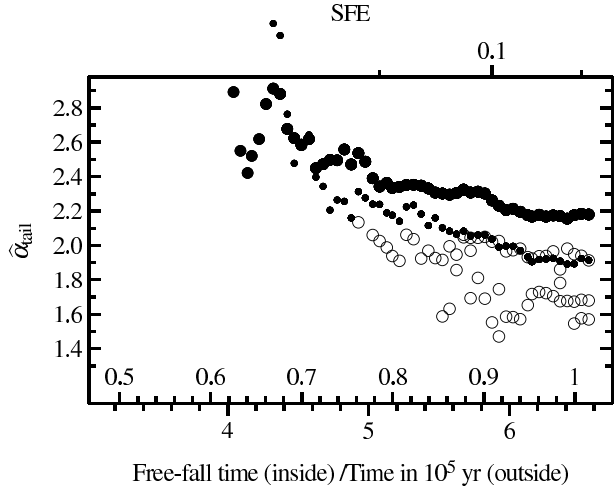}}
\end{center}
\caption{Exponent ($\alpha$) of the upper IMF for individual clusters
(open circles) for the whole simulation (large dots) and for all stars
in clusters (small dots) as a function of time.  (Based on the
simulations of Bonnell \ea 2008: see Maschberger \ea 2009)}
\end{figure}

This is, however, not the full story, since there is also a difference
between the IMF within individual clusters and the IMF of the combined
population of all stars in clusters. This implies that clusters are
not simply randomly assembled from the total population of all stars
in clusters.  Kroupa \& Weidner (2003) first pointed out that random
drawing might not be a good cluster-assembly model and argued that the
integrated IMF for a galaxy (the IGIMF) might be steeper than the IMF
in individual clusters. Weidner \& Kroupa (2006) showed that this
would result if the maximum mass of a star in a cluster is a {\it
systematic} function of cluster mass. (We emphasize the term
`systematic', i.e.  one needs a consistent suppression of massive-star
formation in low-mass clusters, which is {\it stronger} than the
merely statistical effect that, on average, the maximum mass in a
randomly drawn sample is an increasing function of the sample size).
In the simulations we see below that this effect indeed results from a
(cluster-mass-dependent) truncation of the upper IMF in individual
clusters.

\subsection{The truncation of the upper IMF in cluster-formation simulations}

The upper IMF in the simulated clusters is best fit by {\it truncated
power laws} for which the truncation mass depends on the cluster mass
(Maschberger \ea 2009).  We stress that these simulations do not
include feedback, so there is no intrinsic mass scale at which
accretion onto a star is halted. Instead, truncation results from the
finite time that has elapsed since the formation of the oldest stars
in the cluster, which limits the maximum stellar mass attained at a
given time.  Richer clusters create deeper gravitational potentials
and, hence, a richer accretion environment, so the truncation mass is
usually larger in richer clusters.  We note that the oldest, most
massive stars that end up in a particular cluster usually formed as a
small-$N$ entity and that this group of stars travels together through
successive mergers. Thus, the upper end of the IMF in the simulated
clusters is characterized by a group of stars just below the
truncation mass that are rather similar in mass.  (We emphasize that
the simulations extend only over $\sim 0.5$ Myr and, hence, correspond
to the deeply embedded phase of star formation: these age differences
would be imperceptible once the stars emerged as optically visible
sources).

\subsection{Mass segregation in young clusters}

\begin{figure}
\parbox[t]{0.48\columnwidth}
{\includegraphics[width=0.45\columnwidth]{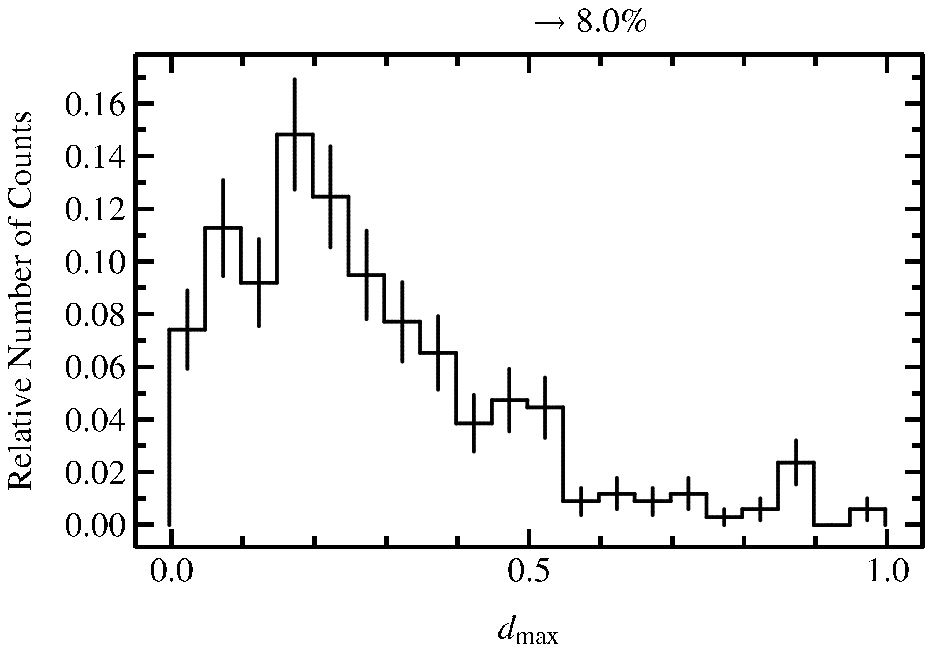}}
\parbox[t]{0.48\columnwidth}
{\includegraphics[width=0.45\columnwidth]{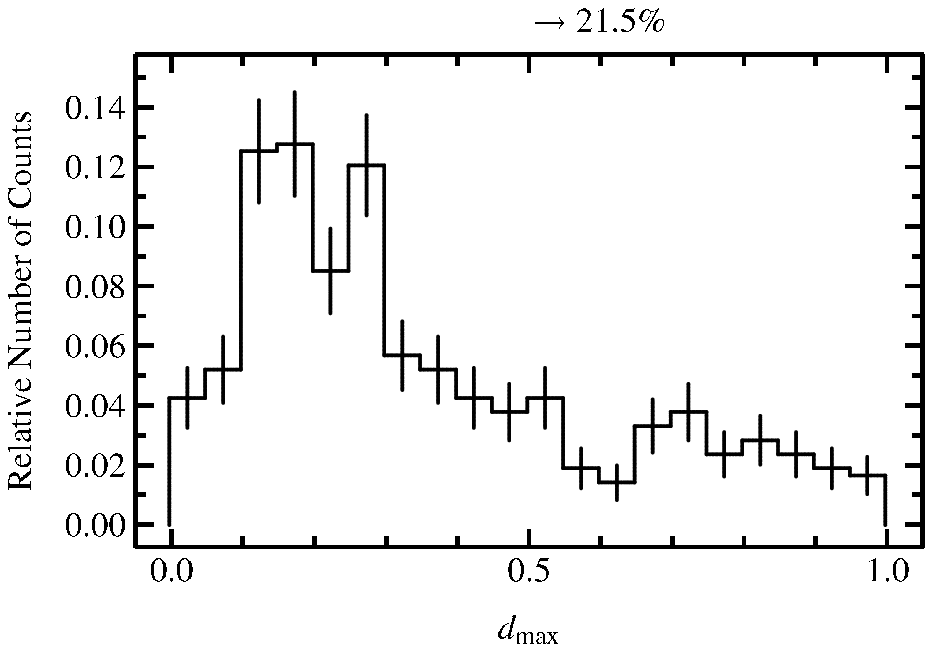}}
\caption{Histogram of fractional radial ranking of the most massive
star in each cluster in the simulations of Bonnell \ea (2003; {\it
left}) and Bonnell \ea (2008; {\it right}: see Maschberger \ea 2009).}
\end{figure}

The simulations show pronounced mass segregation at an age of $\sim
0.5$ Myr. This is demonstrated by figure 6, which shows the
distribution of fractional radial rankings within their parent
clusters of all stars that are the most massive members of their
respective clusters. In the absence of mass segregation, this histogram
would be flat, since the most massive stars would be equally likely to
lie at any radial ranking. Instead, it is clearly peaked towards small
values (the most massive star lies within its parent cluster's
half-number radius 80\% of the time).  Indeed, cases where the most
massive star is significantly offset from its cluster's centre
correspond to situations of ongoing merging.  Figure 6 demonstrates
that periods when mass segregation would not be observed (cf. Bate
2009{\it a}) are relatively brief, implying that the mass-segregation
timescale is generally shorter than the interval between successive
mergers. Hence, mass segregation in clusters of this age should be the
observational norm.

This mass segregation is `primordial' in the sense that it is
established when the cluster is very young by any observational
standards. It is {\it not} primordial, however, in the sense that the
most massive stars did not initially form within the cluster core
(although they may have acquired much of their mass in this
environment).  Instead, the stars that end up with the highest mass
are generally the first to form. Their headstart in mass acquisition
means that they are likely to be deposited in the cluster core in each
successive merger event.  Mass segregation is thus a dynamical
process, although the presence of gas and the highly nonequilibrium
conditions mean that it is not well modelled as a purely two-body
relaxation effect (see also Allison \ea 2009; de Grijs 2010).

\subsection{Cluster morphology}

\begin{figure}
\parbox[t]{0.48\columnwidth}
{\includegraphics[width=0.45\columnwidth]{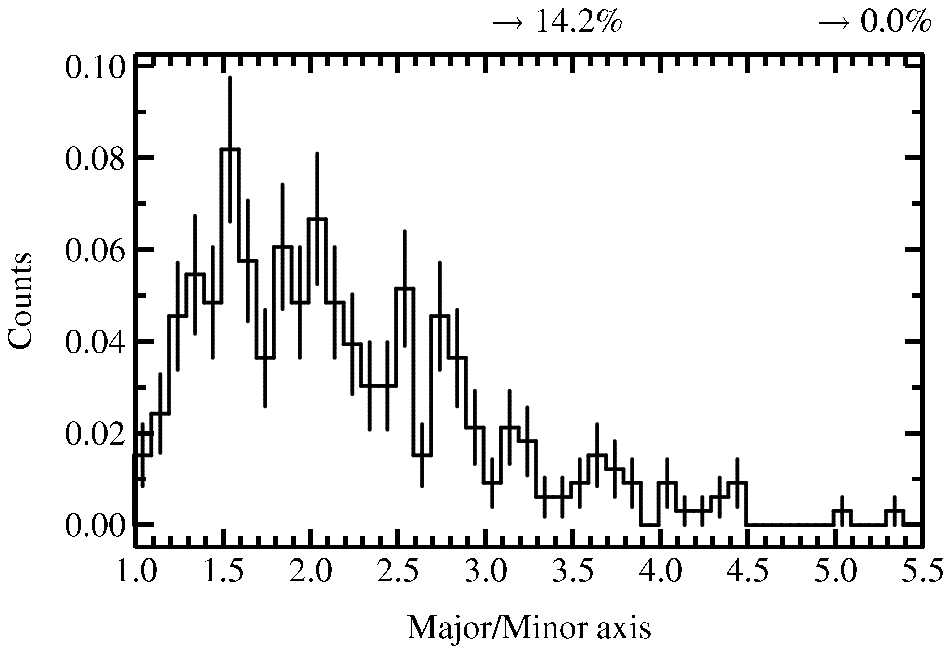}}
\parbox[t]{0.48\columnwidth}
{\includegraphics[width=0.45\columnwidth]{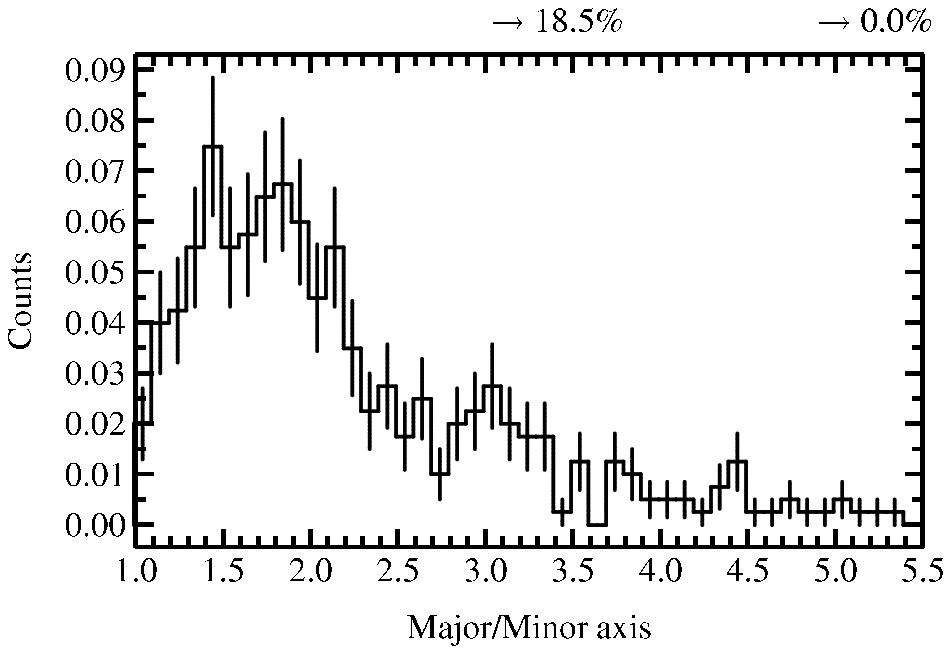}}
\caption{Distribution of projected cluster ellipticities in the
simulations of Bonnell \ea (2003; {\it left}) and Bonnell \ea (2008;
{\it right}: see Maschberger \ea 2009).}
\end{figure}

The projected ellipticities of clusters peak in the range $1-2$
(figure 7). Clusters are thus mildly aspherical, which is a
consequence of two competing tendencies: clusters form from
condensations within filaments and cluster mergers often follow the
filament morphology.  Against this, however, stellar dynamical
interactions in the merging cores tend to sphericalize the
clusters. In general, the cluster cores are more dynamically relaxed
(rounder) than their outer regions, which retain a greater memory of
previous merger events.

\subsection{Dependence of stellar/cluster properties on initial conditions}

Several studies have experimented with changing the power spectrum of
the `turbulent' velocity field, the Mach number of the turbulence and
the global gravitational boundedness (i.e., the ratio of kinetic plus
thermal energy to gravitational energy) of the initial cloud.  Broadly
speaking, the effects of these properties are minor.  The power
spectrum does not significantly affect the clustering statistics of
the resulting stars (Schmeja \& Klessen 2006), as measured, for
example, through the `Q' parameter of Cartwright \& Whitworth
(2004). In the case of one-off turbulent stirring, the power spectrum
also has no effect on the resulting IMF (Bate 2009{\it c}).  If
turbulent forcing is sustained, Klessen (2001) found that low-mass
star formation is suppressed (i.e., the IMF is flattened) when the
turbulent driving is driven on small scales (i.e., comparable with the
Jeans length). As noted above, the turbulent Mach number has little
effect on the resulting IMF (Bonnell \ea 2006) and also does not
affect the clustering statistics (Schmeja \& Klessen 2006).

The gravitational boundedness of the simulation does, however, affect
the degree to which the cluster merger goes to completion (contrast
the left- and right-hand portions of figure 2) and has a striking
influence on the {\it star-formation efficiency} (fraction of cloud
turning into stars per free-fall time; Clark \& Bonnell 2004; Clark
\ea 2008: see \S5$\,c\,$(ii). The effect of boundedness on the IMF is
not clearly established (Clark \ea 2008; Maschberger \ea 2009).

\section{A new generation of simulations: adding extra physics}

The simulations described above have several obvious deficiencies.
The omission of magnetic fields and the crude treatment of the
(barotropic) equation of state encourages profuse fragmentation of
dense gas, thus overproducing low-mass stars and overpredicting the
importance of dynamical interactions. This is particularly acute in
the best-resolved simulations (Bate \ea 2003; Bate 2009{\it a}).  It
is obviously undesirable that such problems are crudely suppressed by
low resolution, as in the large-scale simulations of Bonnell \ea
(2003, 2008). The recent work of Bate (2009{\it b}) and Price \& Bate
(2009; which incorporate radiative transfer and magnetic fields in
cluster-formation simulations) represents an important development,
whose statistical consequences are still relatively unexplored.

The other drawback of the vanilla calculations is the fact that no
feedback from massive stars (either mechanical---winds---or from
thermal ionization) is included. This means that in unmagnetized
clouds that are globally bound the star-formation efficiency would go
to 100\%, thus overpredicting the Galactic star-formation rate by
orders of magnitude. (Note that low efficiency can be achieved without
feedback for initially unbound clouds: Clark \ea 2008; \S5$\,c$).  The
omission of feedback also provides no mechanism for limiting
individual stellar masses if the simulations were run forward in time
to create arbitrarily large-scale clusters.  Observationally, however,
it is clear that the maximum stellar mass never exceeds $150-200$
M$_\odot$, even in very populous clusters (Weidner \& Kroupa 2004; Oey
\& Clarke 2005).  Finally, the omission of feedback from massive stars
means that the issue of triggered star formation is not addressed.

\subsection{Magnetic fields and radiative feedback}

Bate (2009{\it b}) presented turbulent-fragmentation calculations
where the barotropic equation of state was replaced by evolution of
the thermal state of the gas and solution of the equations of
radiative transfer (in the grey flux-limited diffusion limit;
Whitehouse \ea 2005).  Qualitatively, this allows gas to be
radiatively heated by $p\mathrm{d}V$ work liberated deep in the
potential of each forming protostar. (Such calculations, however, {\it
under\,}estimate this effect, since they only include energy liberated
outside the sink radius (at a few AU) and omit the effect of
thermonuclear and accretion energy released at smaller radius: see
Krumholz \ea 2007).  Even this partial inclusion of radiative feedback
changes the simulation outcome quite dramatically at the low-mass end:
the modest rise in temperature in the vicinity of each forming
low-mass protostar increases the Jeans mass and, hence, suppresses
fragmentation locally.  This then severely suppresses the formation of
brown dwarfs, so that the brown-dwarf-to-star ratio is now $\sim 1:5$,
compared with the `vanilla' (barotropic) calculation that yielded a
ratio of $ > 3:2$, in conflict with observations, as discussed
above. Note that the gentle thermal feedback from low-mass star
formation mainly suppresses new object creation. It neither destroys
star-forming cores nor stops accretion onto existing objects.

\begin{figure}
\begin{center}
\includegraphics[width=0.4\columnwidth]{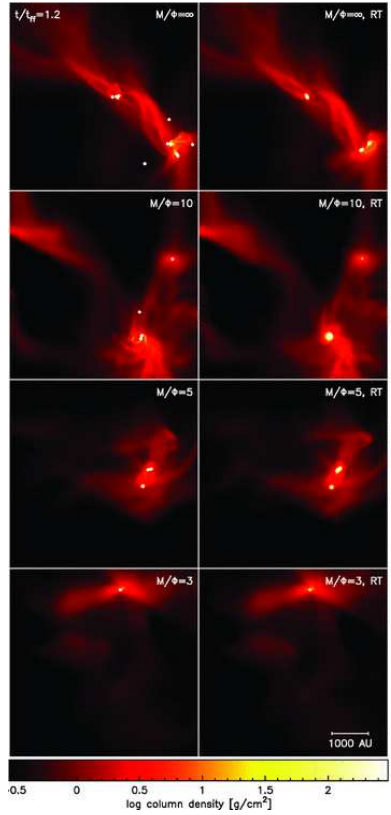}
\end{center}
\caption{Influence of magnetic fields on cloud evolution, showing
suppression of core formation for stronger fields ({\it lower panels}:
Price \& Bate 2009).}
\end{figure}

Price \& Bate (2009) performed the first turbulent-fragmentation
calculations including a (frozen-in) magnetic field, whose initial
amplitude corresponded to a global mass-to-flux ratio of either three
or five times that required for collapse. The magnetic field mainly
affects the large-scale structure by supporting low-density regions
against collapse and thus inhibiting core formation (contrast the
panels in figure 8, which show successively less structure as the
magnetic field is increased down the page).  This reduces the
star-formation efficiency so that $< 10$\% of the cloud is turned into
stars per free-fall time.  This star-formation efficiency is in good
agreement with observational estimates (e.g., Evans \ea 2009), in
contrast to field-free calculations (see figure 10).

\subsection{Thermal ionization and stellar winds}

OB stars feed back energy into the cluster environment both through
the thermal effects of ionizing radiation and via mechanical energy
input from energetic winds. In recent years, each of these effects
have been incorporated into SPH codes [see Kessel--Deynet \& Burkert
(2000), Dale \ea (2007{\it b}) and Gritschneder \ea (2009{\it a}) for
the case of ionizing radiation and Dale \& Bonnell (2008) for winds].

The impact of ionizing-radiation feedback depends quite sensitively on
the location of the ionizing source within the cloud.  In the absence
of feedback, massive stars tend to be located in cluster cores, at the
interception of massive, filamentary accretion flows. The gas
distribution seen from such a star is thus highly anisotropic and this
significantly affects the impact of its ionizing radiation. Dale \ea
(2005) found that the dense filaments remain largely neutral and
inflowing, while low-density regions are rapidly ionized and drive
powerful outflows.  This combined inflow--outflow mode means that
clouds can remain bound even when they absorb (thermal and kinetic)
energy that exceeds their gravitational binding energy. Energy is
carried off by a relatively small mass fraction at speeds that
comfortably exceed the cluster's escape velocity. The highly
inhomogeneous nature of molecular clouds may thus impede their
destruction by ionizing radiation from embedded sources.

\begin{figure}
\begin{center}
\includegraphics[angle=-90,width=0.75\columnwidth]{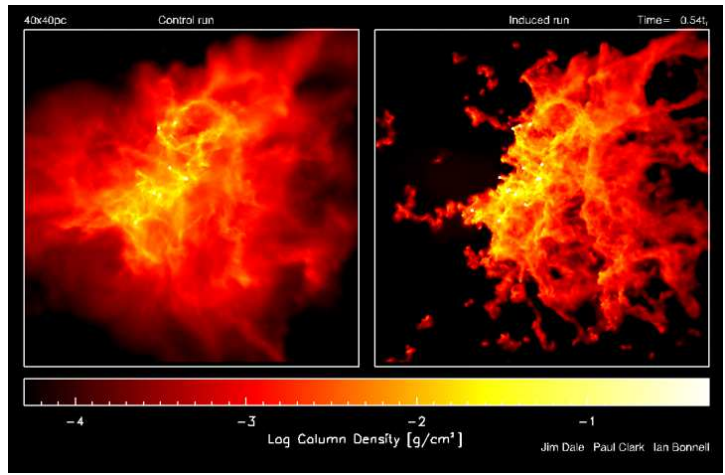}
\end{center}
\caption{Comparison of cloud structure in the presence of an ionizing
source (located to the left of the cloud in the right-hand panel) with
a control simulation {\it (left)}.  (Dale \ea 2007{\it a}.)}
\end{figure}

Dale \ea (2007{\it a}) instead examined the case where the ionizing
source was located {\it outside} a molecular cloud.  This simulation
was globally unbound and so in the control simulation (no ionizing
source) the star-formation efficiency was, in any case, low, with much
of the cloud simply streaming away from a central star-forming core.
When a source of ionizing radiation is introduced, it reverses this
outflow on the side of the cloud facing the source and returns the
material to the dense cloud core, thus boosting the number of stars
formed. This is an example of triggered star formation, although the
result of net positive feedback {\it may} be specific to the situation
modelled of an initially unbound cloud.  Rather disappointingly, no
observational properties were found that could distinguish the extra
stars that formed as a result of ionization triggering from the stars
that formed in the control simulation: both populations were formed in
the dense cluster core and with kinematics dominated by the residual
turbulent motions in the gas.

Dale \& Bonnell (2008) recently included stellar winds in
turbulent-fragmentation calculations, switching on winds (either
isotropic or collimated) with a mass-dependent mechanical luminosity
when stars exceed 10 M$_\odot$.  The winds mainly affect the accretion
rate onto the wind sources themselves and have little effect on either
mass acquisition by lower-mass stars or on the global disruption of
the cloud (it is currently unclear whether or not winds play an
important role in limiting the maximum stellar mass to its observed
value of $150-200$ M$_\odot$). The failure of winds to achieve
complete disruption of the parent cloud (which might be expected given
their high mechanical luminosities) can be traced, as in the case of
photoionization discussed above, to the highly inhomogeneous
conditions in the cloud. The winds effect efficient energy transfer to
the surrounding medium only in the case of low-density regions: these
develop powerful outflows of entrained material but do not disrupt the
inflow in adjacent dense regions.  Although the present pilot
simulations are far from being a comprehensive parameter study, these
early results perhaps indicate a generic problem, i.e., that (in the
case of gravitationally bound clouds with no magnetic field) the
initial collapse may be so rapid that it is hard for feedback to
reverse the dense anisotropic accretion flows that result.

Finally, we note that the inhomogeneity and complex velocity fields in
the simulations mean that the effects of feedback are less visible
than in smooth, static media. Although control simulations show that
clouds are indeed sculpted by both winds and photoionization, these do
not generally produce the spherical bubble features with which such
feedback is conventionally associated.

\section{Conclusions}

\subsection{Observable properties of young star clusters predicted by 
simulations}

There are generic similarities running through all simulations
described above, which make clear predictions about the observational
properties of star clusters at an age of $\sim 0.5$ Myr.  The assembly
of clusters through scale-free hierarchical merging imparts the
resulting clustering with a fractal character, in the sense that the
structures have no preferred mass scale and the pattern of nested
substructures looks similar at all scales.  These clusters are
generally mass segregated, except if they are observed shortly after a
cluster merger event. The time between such events is, however, longer
than the internal mass-segregation timescale, so that mass segregation
is the predicted observational norm.  The clusters are typically
elliptical, their relatively mild flattening (typical axis ratio
$<2:1$) being a trade-off between extension in the plane of the most
recent merger and ongoing sphericalization by relaxation effects.

\subsection{Implications of cluster-formation simulations for 
galactic-scale star formation}

\subsubsection{The IMF in the field versus clusters}

Most of the clusters formed in these simulations are not long-lived
and are expected to contribute to the galactic-field population.  The
largest clusters are $\sim 1000$ M$_\odot$ in mass and thus, depending
on the subsequent history of gas removal from such objects, they might
later be identified observationally as open clusters.  Thus,
simulations can start to shed light on the formation of both the
`field' and `cluster' population in the Galaxy.

In the bottom-up scenario outlined here, essentially all stars are in
clusters at some point.  Since it is the most populous clusters that
are most likely to survive, the ultimate destiny of a particular star
as a field or cluster member is largely determined by how many mergers
it undergoes, which depends at least partly on the the degree of
gravitational boundedness of the parent cloud locally.

Since `cluster' and `field' stars follow similar early evolutionary
histories, one does not expect major differences between `field' and
`cluster' stars. For example, the closest dynamical interactions
(which may determine the formation of binary systems and the possible
destruction of protoplanetary discs) often occur in the early
formation stages when essentially {\it all} stars (regardless of their
ultimate designation as `field' or `cluster' objects) are in small-$N$
groupings. The lower end of the IMF (as measured by the ratio of stars
to brown dwarfs, for example) is thus expected to be similar in rich
clusters and what will ultimately become the field.  In broad terms,
successive cluster mergers do not greatly affect the lowest-mass
stars, since these tend not to undergo significant accretion in
cluster cores.

The upper end of the IMF {\it does}, however, depend on the history of
successive cluster mergers, since mergers furnish the opportunity for
further accretional growth by the most massive stars.  Two
observational consequences of this scenario are therefore (i) that the
slope of the IMF is somewhat flatter in populous clusters than in the
total population and (ii) that the IMF in a cluster is truncated at
the upper end at a value that depends on cluster richness. Weidner \&
Kroupa (2005; see also Pflamm--Altenburg \ea 2009) have set out a
number of potential observational consequences of a scenario in which
the integrated galactic IMF is, in fact, steeper than the IMF in
individual clusters and may vary with galactic properties [see
Hoversten \& Glazebrook (2008) for possible observational
corroboration of this effect].  The simulations support such a
distinction, but in the cases studied so far the difference is rather
modest (see figure 5).

Finally, it should be stressed that the above is derived from
simulations that omit stellar feedback. Pilot studies involving both
ionizing radiation and feedback from stellar winds suggest that the
latter may be important in limiting individual stellar masses at the
high-mass end and may thus introduce an additional truncation to the
stellar IMF that is independent of cluster environment.

\subsubsection{The efficiency of star formation}

\begin{figure}
\parbox[t]{0.48\columnwidth}
{\includegraphics[width=0.45\columnwidth]{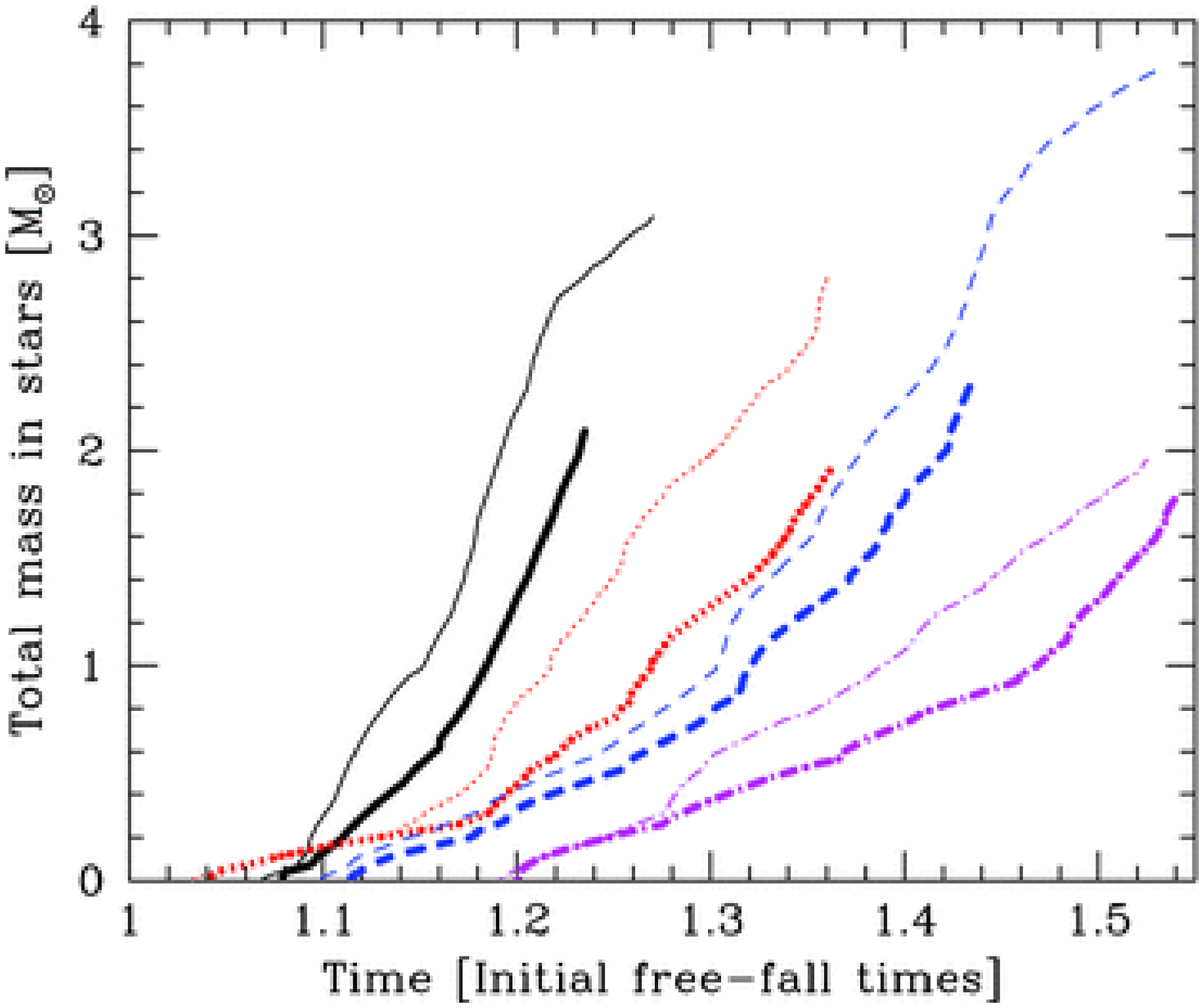}}
\parbox[t]{0.52\columnwidth}
{\includegraphics[width=0.435\columnwidth]{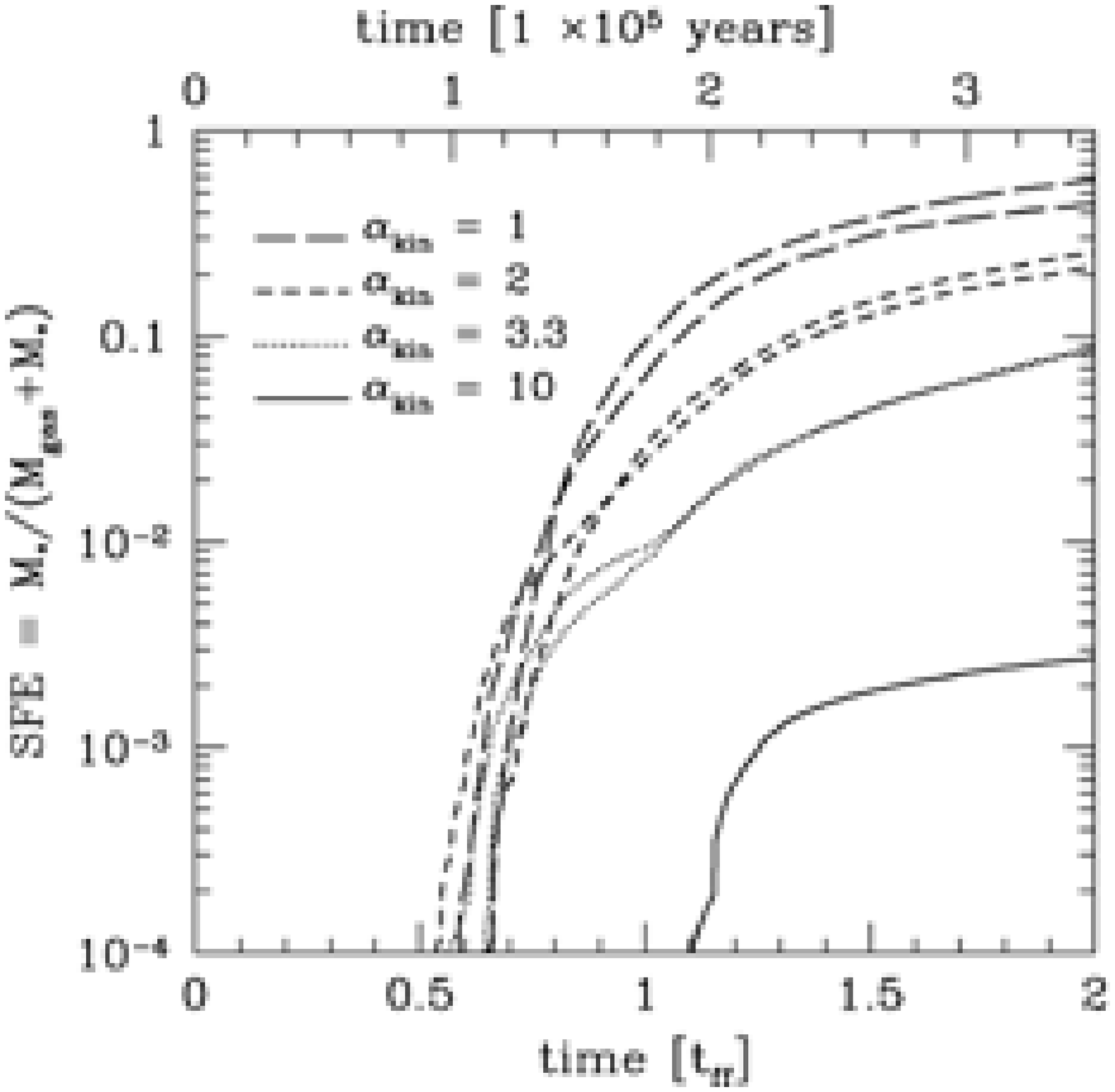}}
\caption{Effect of magnetic-field strength (increasing left to right
in the left-hand panel; Price \& Bate 2009) and initial ratio of
kinetic to gravitational energy (increasing left to right in the
right-hand panel; Clark \ea 2008) on star-formation efficiency.}
\end{figure}

The fraction of a cloud that is incorporated in stars per free-fall
time is an important quantity since it controls the relationship
between the total molecular gas in the Galaxy and the Galactic
star-formation rate. Current observational estimates of these
quantities imply a star-formation efficiency of $<10$\% (Evans \ea
2009). It is found that, in the absence of magnetic fields,
simulations starting from gravitationally bound initial conditions
form stars much more rapidly than this. In fact, pilot simulations
(Dale \ea 2005; Dale \& Bonnell 2008) suggest that, in this case, the
collapse is so rapid that by the stage that feedback from massive
stars (photoionization or winds) switches on, it can be hard to
disrupt the dense accretion flows and quench the star-formation rate
significantly.

Instead, it would seem that low star-formation efficiencies can be
achieved either in simulations that start from globally unbound clouds
(Clark \& Bonnell 2004; Clark \ea 2008) or where clouds are threaded
by (moderately supercritical) magnetic fields (Price \& Bate 2009):
see figure 10.  Either of these conditions helps to stop runaway
collapse in the early stages of cloud evolution and thus may make it
easier for stellar feedback to quench star formation at a later
stage. This is, however, speculation: currently, simulations have just
reached the stage where they have started to implement these effects
separately.  Further investigation is required to understand how these
processes interact in the formation of stars and clusters from
molecular clouds.

\end{document}